\documentclass[12pt]{article}
\usepackage{amsmath}
\usepackage{epsfig}
\usepackage{graphics}
\usepackage{here}
\begin{document}
\begin{titlepage}
\begin{center}
\vspace*{1.0cm}
\begin{tabbing}
xxxxxxxxxxxxxxxxxxxxxxxxxxxxxxxxxxxxxxxxxxxxxxxxxxxxxxxxxxxx \= \kill
\> {\bf NuMI-B-786} \\
\end{tabbing}
\vskip 2cm

{\LARGE \bf Neutrino Oscillations Experiments using Off-axis NuMI Beam}
\vskip 1cm
    
\vspace*{5mm}
{\it \large \today}\\  
\vspace*{1.0cm}
A. Para, Fermilab

\vspace*{5mm}

M. Szleper, Northwestern University
\end{center}
\vskip 2cm
\begin{abstract}
NuMI neutrino beam is constructed to aim at the MINOS detector in Soudan mine.
Neutrinos emitted at angles $10-20\;mrad$ with respect to the beam axis
create an intense beam with a well defined energy, dependent on the angle.
Additional surface detectors positioned at the transverse distance
of several kilometers from the mine offer an opportunity for  very precise 
mesurements of the neutrino oscillation parameters. The mixing
matrix element $\left|U_{e3}\right|^{2}$ can be measured down to a value 
of 0.0025 with the exposure of the order of $20\;kton\times years$.    
\end{abstract}
\end{titlepage}
\newpage
\newpage
\tableofcontents
\newpage
\listoffigures

\section{Introduction}

A discovery of the neutrino oscillations by the SuperK\cite{SuperK} experiment
has changed the neutrino physics. The existence of the phenomenon is no longer
in question and the focus of the experiments has changed from the 'discovery' 
to the 'elucidation'. The next round of experiments, running or under 
construction, will provide an independent verification of the SuperK result
and it is likely confirm that $\nu_{\mu}$ neutrinos oscillate mostly into
$\nu_{\tau}$ with neutrinos with a characteristic frequency corresponding to
$\Delta m^{2}\sim 1.5-4\times 10^{-3}\,eV^{2}$. 

Questions related to neutrino masses and oscillations are of a fundamental
nature  therefore  a significant  effort is dedicated to a possible design
of the future facilities: high intensity conventional neutrino beams,
often referred to as 'superbeams', and/or a novel facility, 'neutrino factory',
where a very intense mixed beam of $\nu_{\mu}$'s and $\nu_{e}$'s is produced
from decay of muons circulating in a storage ring.

In this note we examine a potential extension of the utilization of the 
NuMI beam, being constructed at Fermilab, by addition of new detectors.
These new detectors could offer a  major enhancement of the physics reach
of the MINOS detector, especially in conjunction with the anticipated
improvements of the Fermilab's accelerator complex.

\section{The NuMI Beams}
\label{numi_beam}

Neutrino beam constructed at Fermilab has a flexible  design\cite{NUMI}. 
Focusing 
elements, consisting of two parabolic horns, can be be moved with respect 
to the target and to each other thus focusing different momentum bytes of the 
produced pions and giving rise to the neutrino beams of different energies.

\begin{figure}[h]
\centerline{\epsfig{file=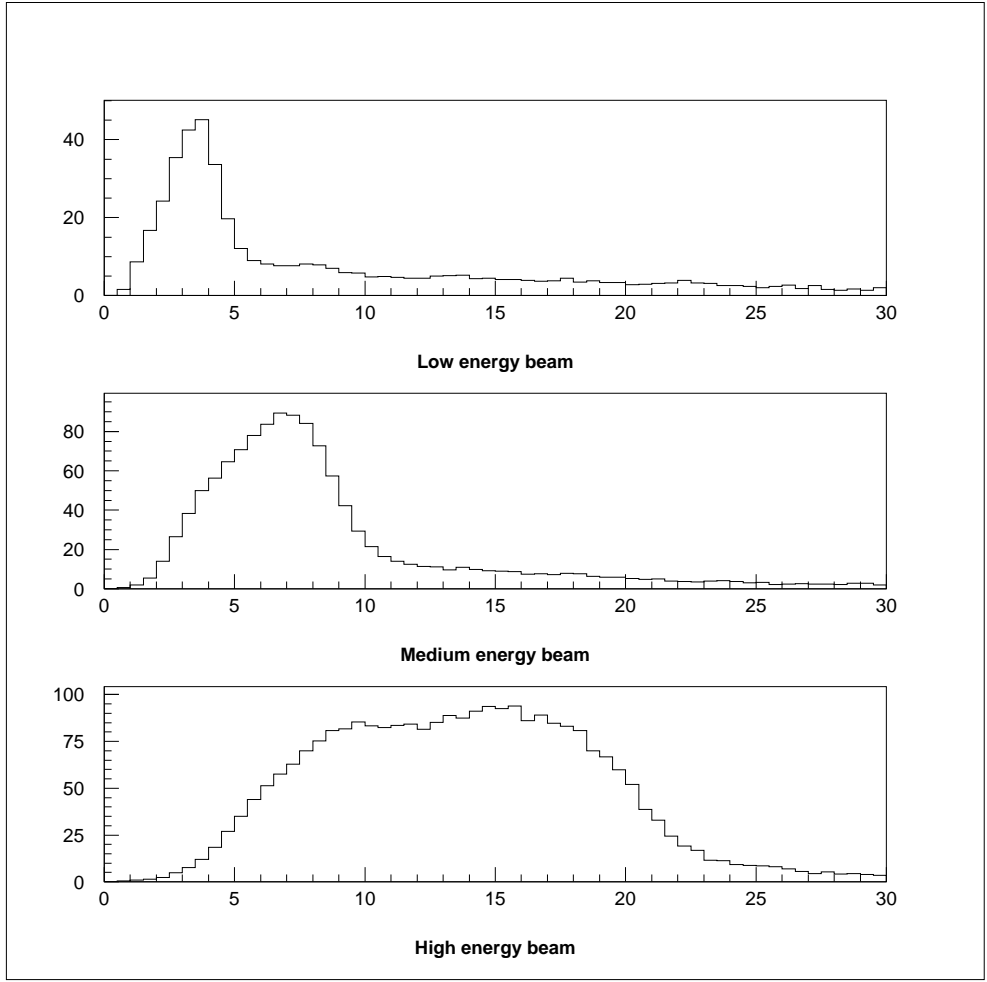,height=13cm}}
\caption{High, medium and low energy options for NuMI beams. Histograms 
represent the rates of $\nu_{\mu}$ CC events expected at the distance of
$730\,km$ in a detector with $1\,kton$ fiducial volume.}
\label{NuMI}
\end{figure}

The resulting beams are presented in Fig.\ref{NuMI} showing the event spectra
for a nominal $1\, kton\times year$ exposure at the distance of 730 km from the target.
In this  paper we assume the proton flux on target to be $3.8\times10^{20}$
per year.

The horn acceptance is defined in terms of the relative momentum spread
$\frac{\Delta p}{p}$ producing the neutrino beam with the energy spread
increasing with the beam energy. The long tail of the high energy neutrinos 
in the low and medium energy beams is due to the bare target component
of the beam created by 
high momentum pions traversing the opening in the magnetic horns.    

\section{The Principle of Off-axis Neutrino Beams}
\label{off-axis}

Most of the neutrinos in a conventional neutrino beam are
produced in two-body decays $\pi^{+} \rightarrow \mu^{+} + \nu_{\mu}$.
A smaller contribution ti the beam is due to $K_{\mu3}$ decays,
 $K^{+} \rightarrow \mu^{+} + \nu_{\mu}$, and  to the decays of the
muons, $\mu^{+} \rightarrow e^{+} {\overline{\nu_{\mu}}} \nu_{e}$.

Energy and the flux of the neutrinos produced in two-body decays 
is uniquely determined by the decay  angle $\theta_{dec}$. For  small
angles they can be expressed as:
\begin{equation}
  E_\nu=\frac{0.43E_{\pi}}{1+\gamma^{2}\theta^{2}_{dec}}  
\end{equation}

\begin{equation}
Flux= \left(  \frac{2\gamma}{1+\gamma^{2}\theta^{2}_{dec}} \right) ^{2}
\frac{A}{4\pi z^2}
\label{flux}
\end{equation}

where 
\begin{itemize}
\item $\gamma=\frac{E_\pi}{m_{\pi}}$ is the Lorentz boost factor of a pion
\item $\theta_{dec}$  is decay angle, i.e. the angle between the pion and the
produced neutrino directions
\item $A$ is the area of the detector and $z$ is the distance between the decay
point and the detector.
\end{itemize}

\begin{figure}[h]
\centerline{\includegraphics*[bb=30 160 550 690,height=12cm]{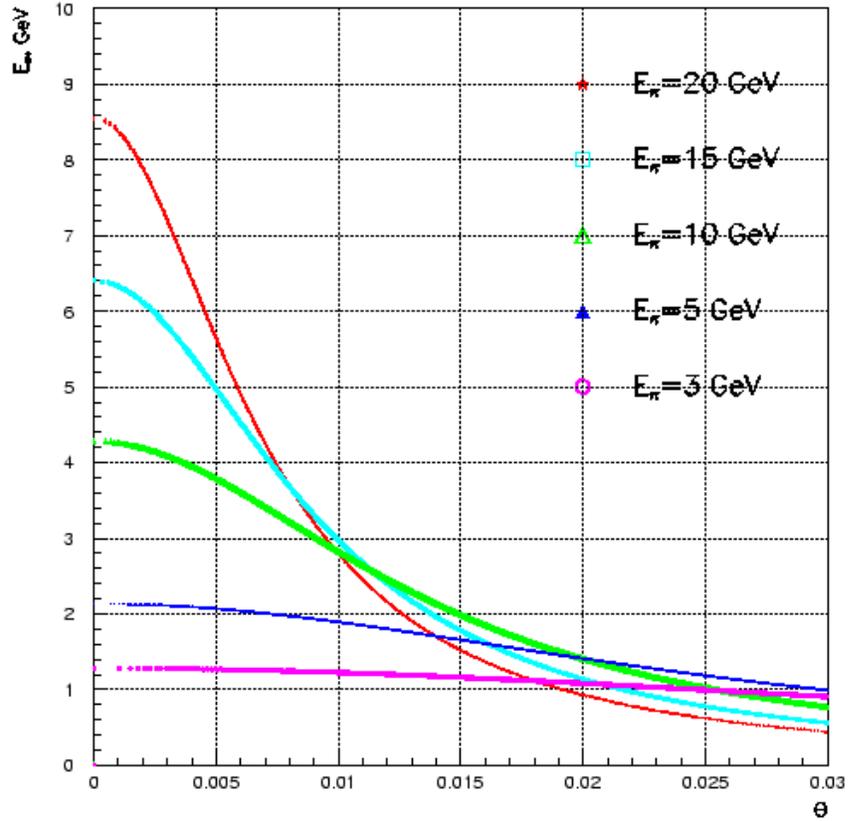}}
\caption{Variation the neutrino energy with the decay angle $\theta_{dec}$
for neutrinos produced by $\pi$'s of different energies.}
\label{e_vs_theta}
\end{figure}

\begin{figure}[h]
\centerline{\includegraphics*[bb=30 160 550 690,height=12cm]{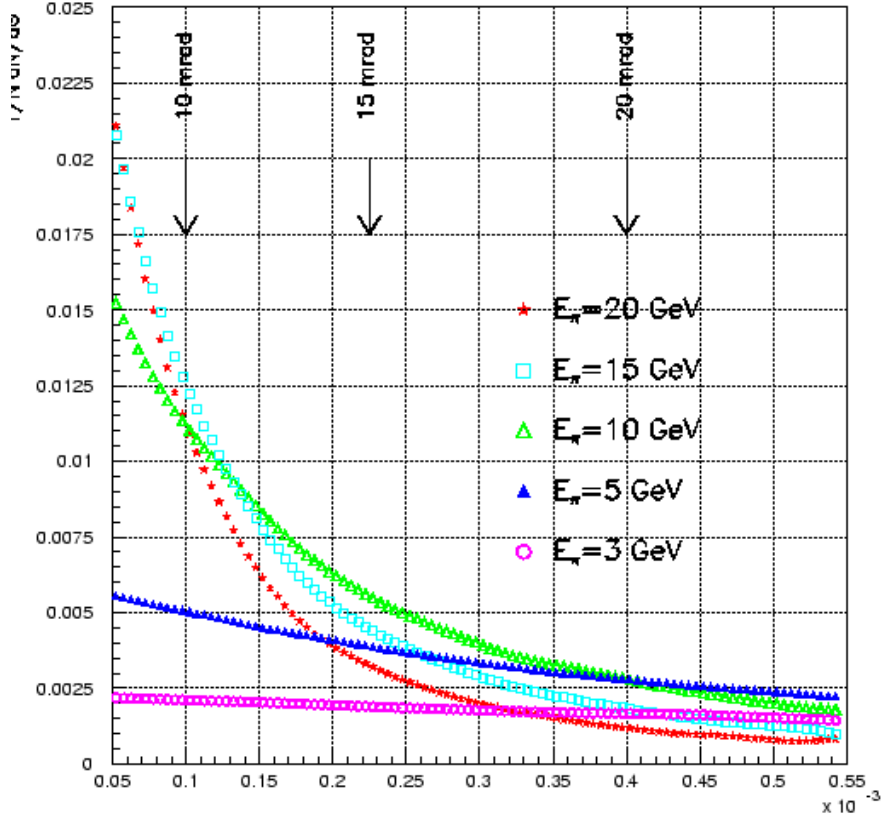}}
\caption{Variation the neutrino flux with the decay angle $\theta_{dec}$
for neutrinos produced by $\pi$'s of different energies.}
\label{flux_vs_theta}
\end{figure}

Figs.\ref{e_vs_theta} and \ref{flux_vs_theta} illustrate the relationship between
the neutrino fluxes, energies and the decay angles for the pions in the
energy ranges relevant to the NuMI beam. Neutrino flux is maximal in the 
forward direction, $\theta_{dec}=0$, and the neutrino energy there is proportional
to the energy of the parent pion, $E_{\nu}=0.43E_{\pi}$. This is the
reason for placing the neutrino detector 'on-axis'.

Fig.\ref{flux_vs_theta} shows that the sharp peak of the neutrino flux
in the forward direction changes into a broad maximum for pions
at low energies, $E_\pi \leq 5\,GeV$. Low energy neutrino beam, 
$E_{\nu}\leq 2-3\,GeV$,
is very broad  and the detector location is not very critical
for the energy and intensity of the neutrino flux. This fact makes 
low energy neutrino oscillation experiments relatively insensitive to the
details of understanding of the underlying neutrino beam.

The same fact enables a construction of improved, with higher intensity
and smaller energy spread, low energy neutrino beams. This technique, 
pioneered in the Brookhaven neutrino oscillation experiment proposal\cite{BNL},
consists of placing a neutrino detector at some angle with respect to the
conventional neutrino beam. 

Such a detector records approximately the same flux of low energy neutrinos,
as the one positioned 'on-axis', originating from the decays of low energy pions.
In addition, though, an 'off-axis' detector records an additional contribution
of low energy neutrinos from the decay of higher energy pions decaying at the
finite angle $\theta_{dec}$. 
  
An additional attractive feature of the neutrino flux observed at the
'off-axis' detector is a kinematical suppression of high energy neutrino
component. Detectors placed at different angles with respect to the
neutrino beam direction are therefore exposed to an intense  narrow-band 
neutrino with the energy defined by the detector position.

In contrast with the traditional narrow-band neutrino beams the off-axis
detectors enable a simultaneous experiments with beams at different 
energies/locations.  At the same time the `off-axis' neutrino flux is much higher 
than the one achievable with the traditional narrow-band beam.

\section{Off-axis Beams at NuMI}

Neutrino beam described in the previous section is directed towards
a MINOS detector located in the Soudan mine. This beam, especially in its 
low- and medium-energy
version is very broad and its energy spectrum changes significantly
with the position/angle of the neutrino detector.

We consider a hypothetical situation, where the 'on-axis' MINOS detector
is complemented by three additional detectors positioned at the 
transverse distance of 5, 10 and 20 km away from the central detector.
For a sake of an example we examine a case of $10\,kton\times year$ exposure.

\subsection{Low Energy Beam Case}  

\begin{figure}[h]
\centerline{\epsfig{file=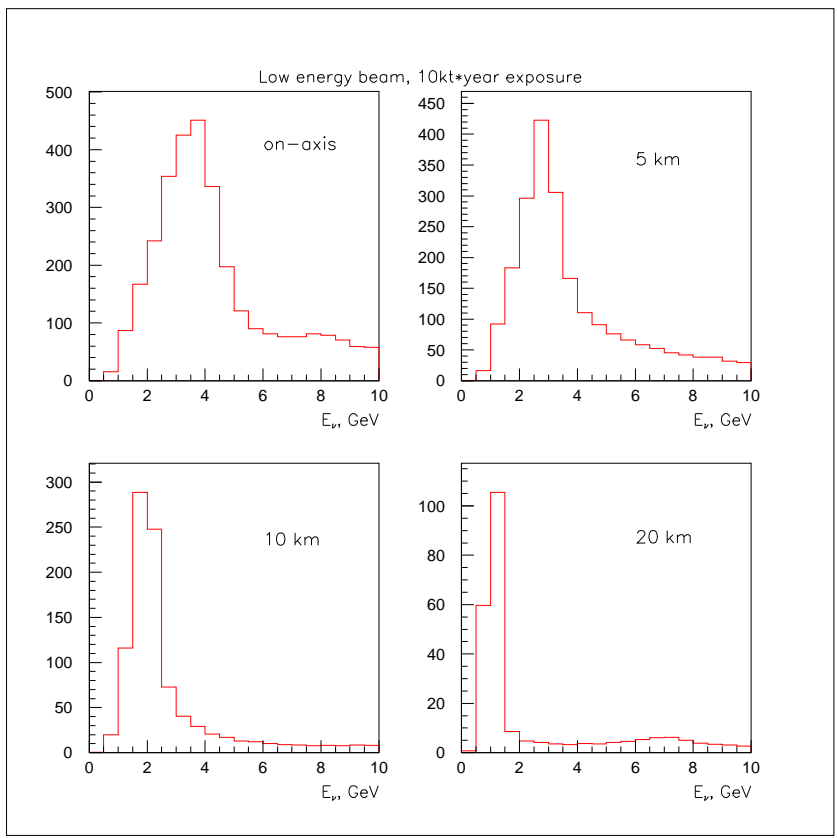,height=11cm}}
\caption{Low energy beam option: neutrino event spectra for 10 kton*years 
exposure
at the detectors on-axis and at 5,10 and 20 km distance.}
\label{le_off_axis}
\end{figure}

Fig.\ref{le_off_axis} shows the expected (in 'no oscillations') case rates
of $\nu_{\mu}$ CC events in the four detectors.

\begin{figure}[h]
\centerline{\includegraphics*[bb=30 160 550 665,height=13cm]{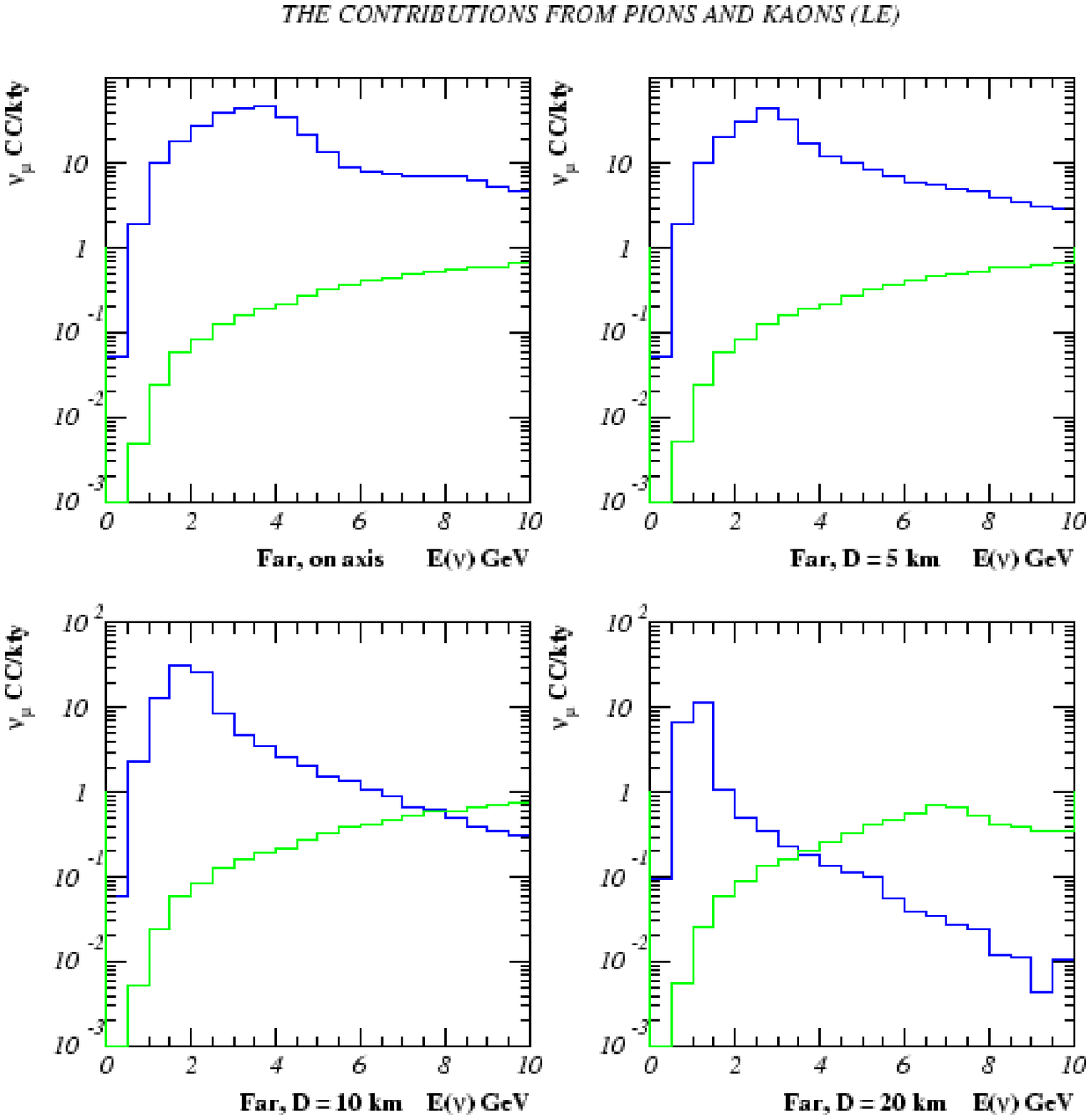}}
\caption{Contribution of pion- and kaon-produced neutrinos to the neutrino
flux at the detectors on-axis and at 5,10 and 20 km distance, low energy beam.}
\label{le_pik}
\end{figure}

As expected, the neutrino spectrum is much narrower at the 'off-axis'
detectors than the one at the central detector, with the peak position
shifting to the lower energies with the increasing distance from the beam axis.
The high(er) energy tail of the spectrum is due to high momentum pions
produced at relatively large angles and decaying shortly after the production
target and due to kaon decays, as shown in Fig.\ref{le_pik}.

\subsection{Medium Energy Beam Case}

\begin{figure}[h]
\centerline{\epsfig{file=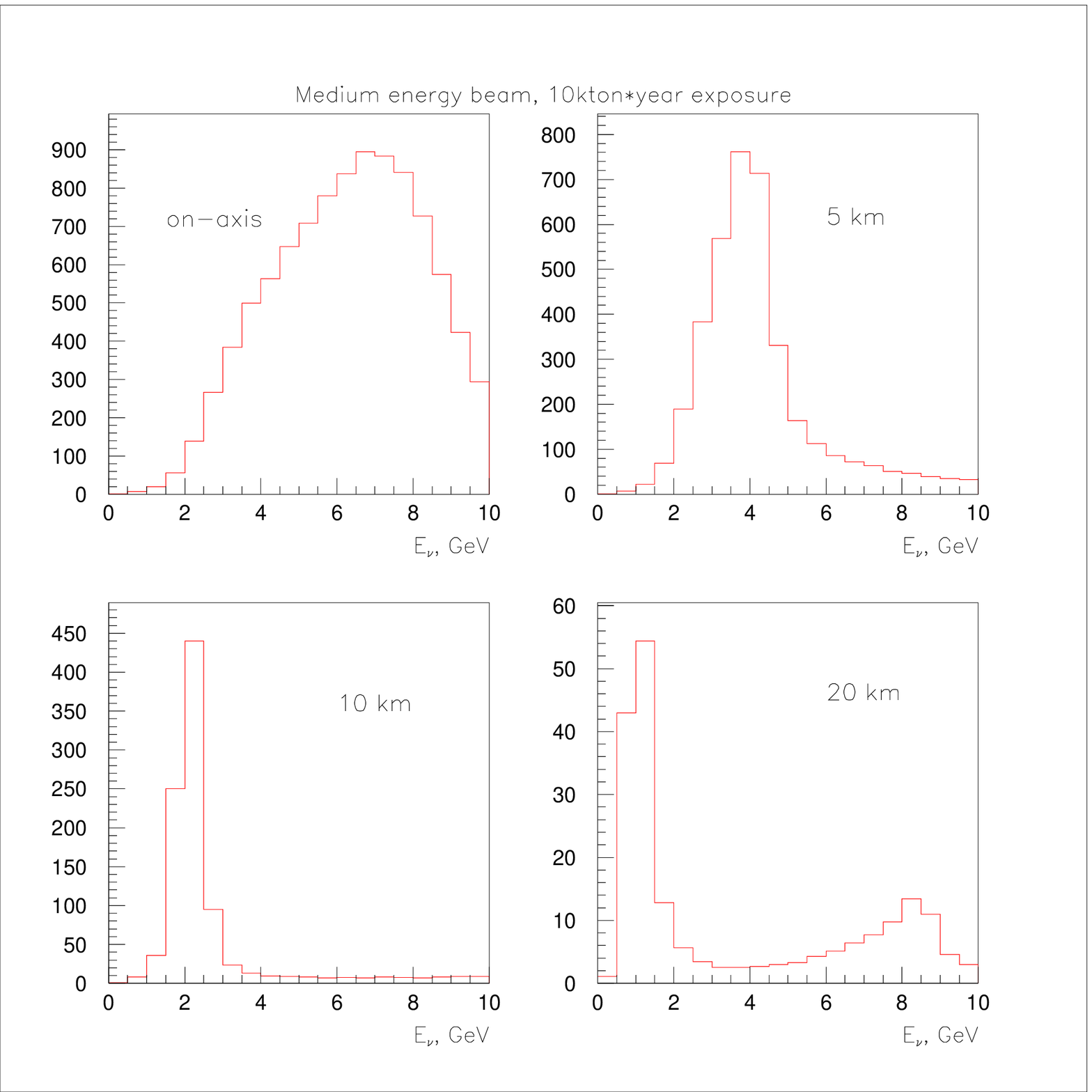,height=11cm}}
\caption{Medium energy beam option: neutrino event spectra for 10 kton*years 
exposure at the detectors on-axis and at 5,10 and 20 km distance.}
\label{me_off_axis}
\end{figure}

Expected neutrino events spectra  at the 'off-axis' detectors in the case
of the medium energy NuMI beam are shown in 
Fig.\ref{me_off_axis}.

This configuration seems to be close to the optimal running conditions.
The neutrino flux in the $3-4\,GeV$ region
at the distance of $5\,km$ is higher than the peak of the low energy beam
'on-axis', with the high energy tail greatly reduced. At the distance
of $10\,km$ the flux of the
neutrinos in $1-2\,GeV$ region is much higher, whereas the contribution of 
higher energy neutrinos is suppressed in comparison with the same detector
exposed to the low energy beam in 'on-axis' position.
 
\subsection{High Energy Beam Case}

\begin{figure}[h]
\centerline{\epsfig{file=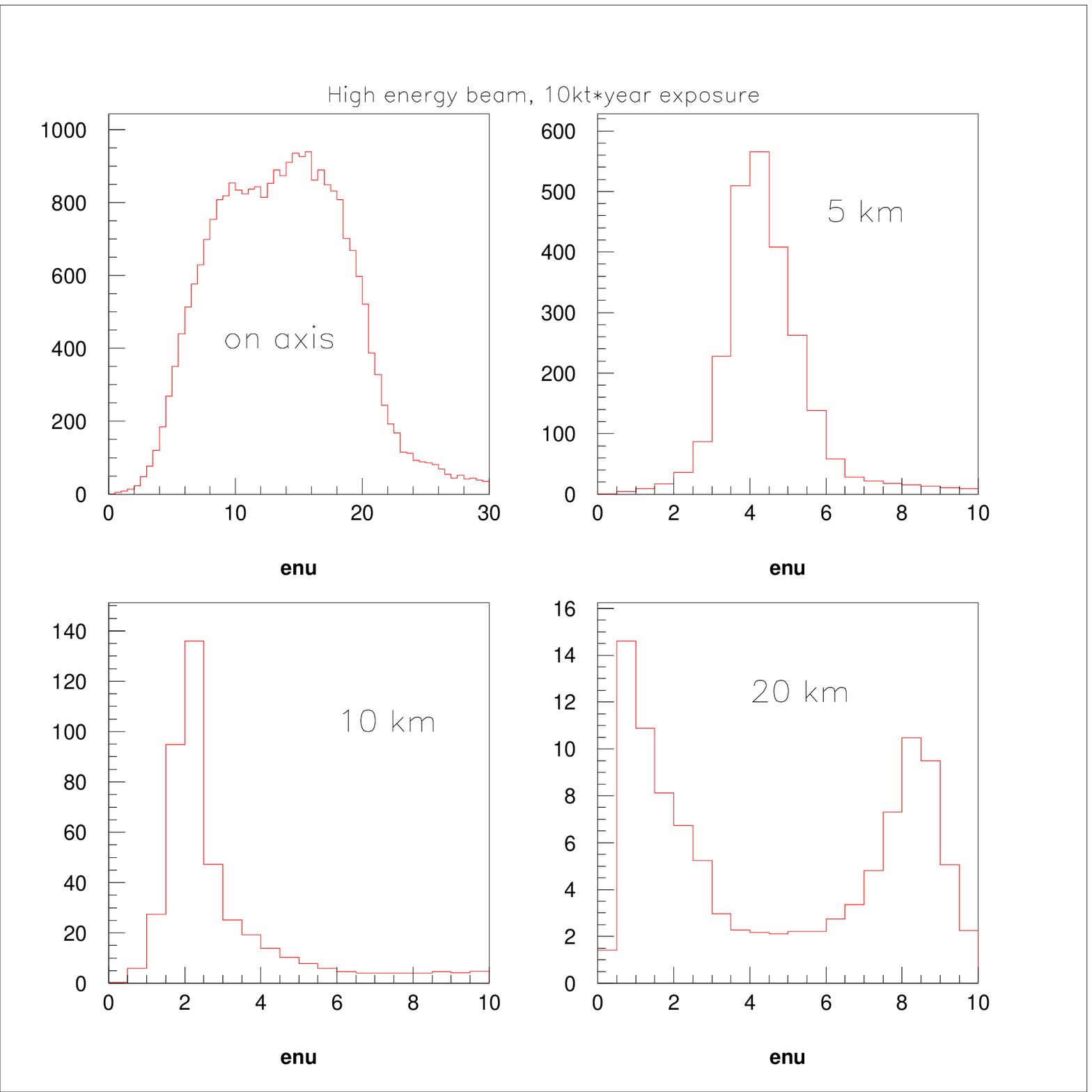,height=11cm}}
\caption{High energy beam option: neutrino event spectra for 10 kton*years 
exposure
at the detectors on-axis and at 5,10 and 20 km distance.}
\label{he_off_axis}
\end{figure}

High energy beam is derived from high energy pions produced in the target.
At these high energies a sharp drop of the neutrino flux intensity, related
primarily to the large factor $\gamma=E_{\pi}/m_{\pi}$ in the Eq.\ref{flux}
leads to the effective beam intensities at the 'off-axis' detectors 
significantly reduced in comparison with the low and medium energy beams,
as shown in Fig.\ref{he_off_axis}

\section{ $\nu_{\mu}$ Disappearance Experiment with the Off-axis Detectors}

A disappearance experiment consisting of several detectors positioned at 
different distances from the neutrino beam axis has several attractive 
features:
\begin{itemize}

\item average energy of the neutrino beam can be selected by the position
of the detectors. A suitable choice of the detectors location will enable
detection of a  oscillation pattern whereby the observed
event rate goes down and up with the increasing distance of the detector
from the beam and hence with the decreasing neutrino energy, as illustrated 
on Fig.\ref{me_dm_030}.

\begin{figure}[h]
\centerline{\includegraphics*[bb=50 160 550 480,height=8cm]{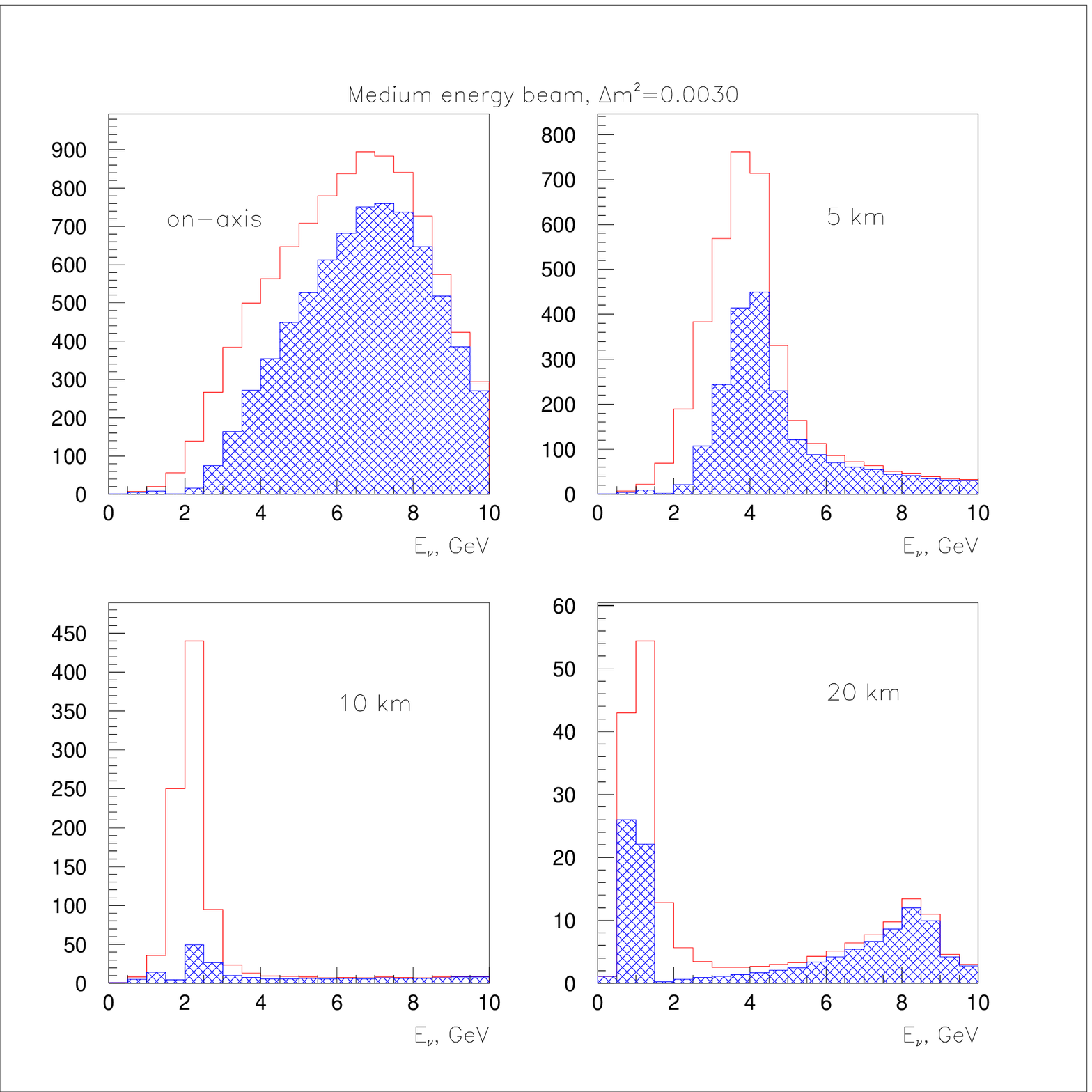}}
\vskip 3.5cm
\caption{Expected $\nu_{\mu}$ CC event rates it the on- and off-axis
detectors for 10 kton*years exposure in the medium energy beam. The histogram
represents 'no-oscillation' case. The cross-hatched distribution shows the 
expected rate in case of oscillations with full mixing 
and $\Delta m^{2}=3.0\times 10^{-3}\,eV^{2}$.}
\label{me_dm_030}
\end{figure}

\item measurement of the disappearance of  $\nu_{\mu}$ CC events can
be easily extended to very low energies,  $E_{\nu} \leq1\,GeV$ . At these
low energies the CC events will be primarily of quasi-elastic type, 
$\nu_{\mu}+n \rightarrow \mu^{-}+p$ and will
consist  of a single penetrating muon track.
In case of a wide band beam in the 'on-axis' detector such events
will be contaminated by single pion events produced by the NC interactions 
of neutrinos at higher energies.

\item a judicious choice of the detector location in conjunction with the
small energy spread of the main peak of the neutrino flux enables a very 
precise  determination of the mixing angle $sin^{2}2\theta_{23}$. As an 
example, in case of $\Delta m^{2}_{23}=3\times 10^{-3}\,GeV$ a 10 kton detector
positioned at the distance of 10 km  and exposed to the neutrino beam
produced by $8\times10^{20}$ protons on target will observe 94 $\nu_{\mu}$ CC events
instead of the expected $821$ events. The number of 'missing' events
provides a direct determination of the mixing angle $sin^{2}2\theta_{23}$
with the  statistical accuracy $\Delta sin^{2}2\theta_{23} \leq 0.0014  $.

\item the average energy of the  neutrinos in the peak of the spectrum is 
determined from two-body kinematics and the detector position, independent
of the detector resolution. This is a significant advantage over the wide-band 
neutrino beam, where the  kinematical effect of the masses of the produced 
particles and nuclear effects introduce systematic difference between
the observed (even with an ideal detector) total energy of the final state
and the energy of the incoming neutrino. This effect is likely to limit the
ultimate  precision of the determination of    $\Delta m^{2}_{23}$ by the
neutrino oscillation experiments, baring a breakthrough in the understanding
of the intra-nuclear cascades and other nuclear effects.

\end{itemize}

\section{Systematics I: Understanding of the Beam Spectra}

Well defined beam energy, determined completely by the detector position
 and a small energy spread make the 'off-axis' beam potentially a very 
powerful tool for studies of neutrino oscillations. For this purpose it 
is necessary, however, to be able to predict the expected neutrino beam,
its spectrum and the  overall normalization with adequate precision.

The far detector spectra in the 'on-axis' can be determined from the event 
spectra observed at the near detector. Does the experiment with 'off-axis'
 detector require a dedicated near experiments near the neutrino source?
Such an experiment would be complicated by the extended nature of the neutrino
source. 

\begin{figure}[h]
\centerline{\includegraphics*[bb=50 160 550 650,width=12cm]{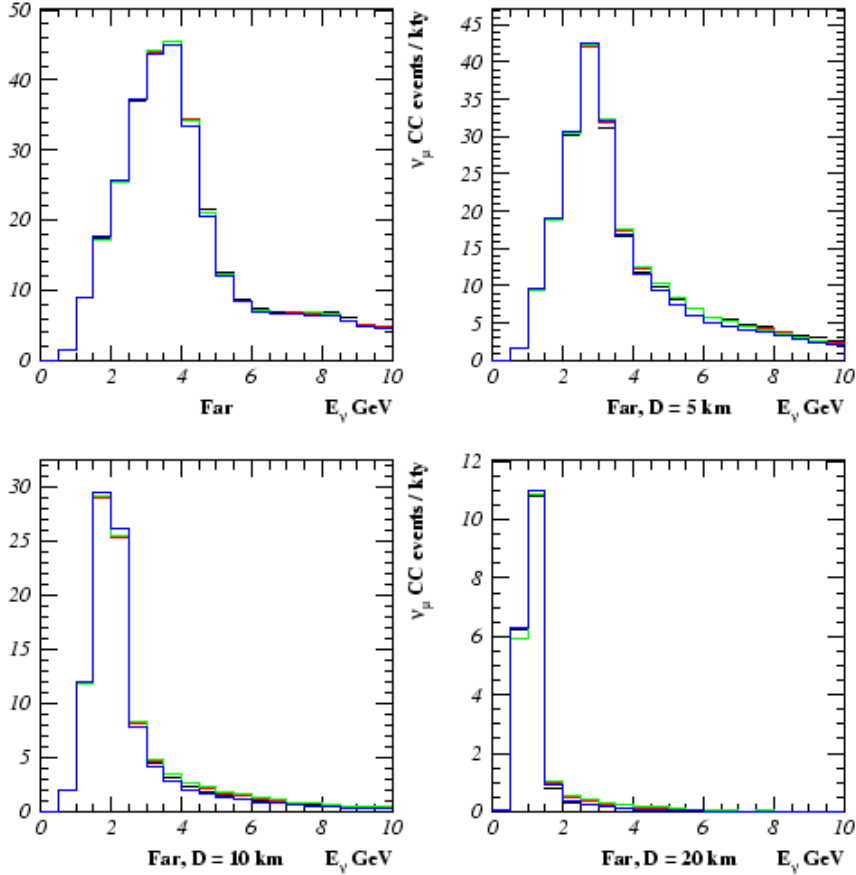}}
\caption{ Dependence of the predicted neutrino spectrum on the hadron
production model for the case of the low energy beam. Four overlayed histograms 
represent the expected spectra at different detector positions derived from the
spectrum observed at the near detector using different hadron production models.}
\label{le_had_prod}
\end{figure}

\begin{figure}[h]
\centerline{\includegraphics*[bb=50 160 550 650,width=12cm]{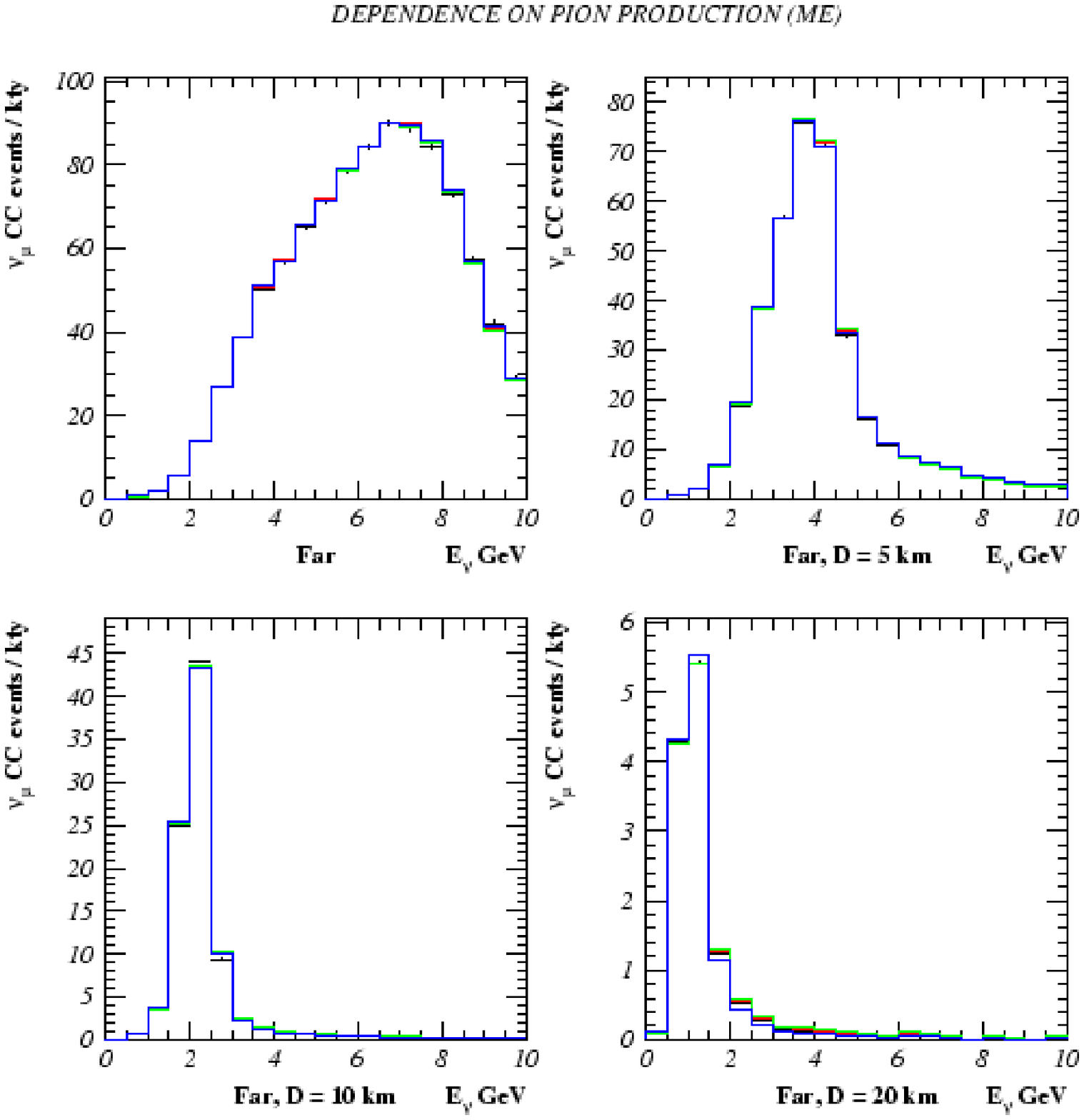}}
\caption{ Dependence of the predicted neutrino spectrum on the hadron
production model for the case of the medium energy beam. Four overlayed histograms 
represent the expected spectra at different detector positions derived from the
spectrum observed at the near detector using different hadron production models.} 
\label{me_had_prod}
\end{figure}

Fortunately, such an experiment is not necessary. The neutrinos observed in the
near detector and in any of the far detectors are derived from the same 
parent beam, formed by the focusing elements. This common origin implies strong
correlations between the beams observed at different locations. The correlation
matrix is primarily determined by the geometry of the neutrino beam and by 
the details of the focusing system, thus permitting an unambiguous prediction
of the neutrino flux at the arbitrary position using the spectrum observed
at the near detector\cite{our_paper}. A residual dependence on the hadron
production model is very small, both for the case of the low and medium energy beams, 
as indicated by the Figs.\ref{le_had_prod} and \ref{me_had_prod}.

\section{Systematics II: Backgrounds}

The principal limitation of a sensitivity of a possible $\nu_{e}$ appearance
experiment will be related to the $\nu_{e}$ and $\overline {\nu_{e}}$ component
of the beam. They are produced in $K^{+} \rightarrow \pi^{0}e^{+}\nu_{e}$
and $\mu^{+} \rightarrow e^{+}\nu_{e}\overline {\nu_{\mu}}$ decays. The
angular distribution of these three-body decays is somewhat different from 
the decays producing the main component of the $\nu_{\mu}$ beam.  

\begin{figure}[h]
\centerline{\includegraphics*[bb=50 160 550 650,width=12cm]{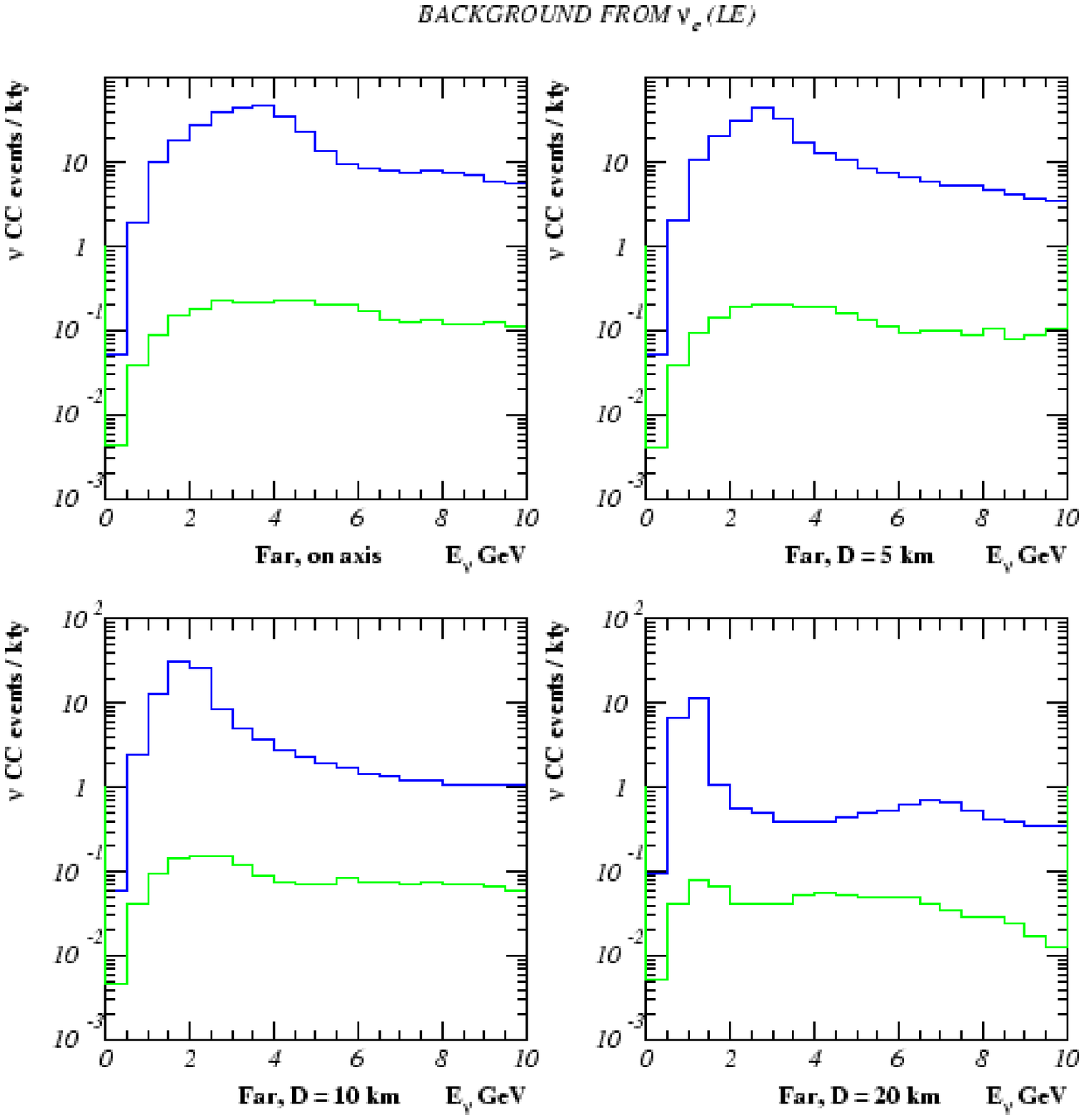}}
\caption{$\nu_{\mu}$ and $\nu_{e}$ induced CC event rates at the
on- and off-axis detectors, low energy beam.The upper histogram represents the
$\nu_{\mu}$ rates, the lower one shows the $\nu_{e}$ rates.}
\label{nue_back_le}
\end{figure}

\begin{figure}[h]
\centerline{\includegraphics*[bb=50 160 550 650,width=12cm]{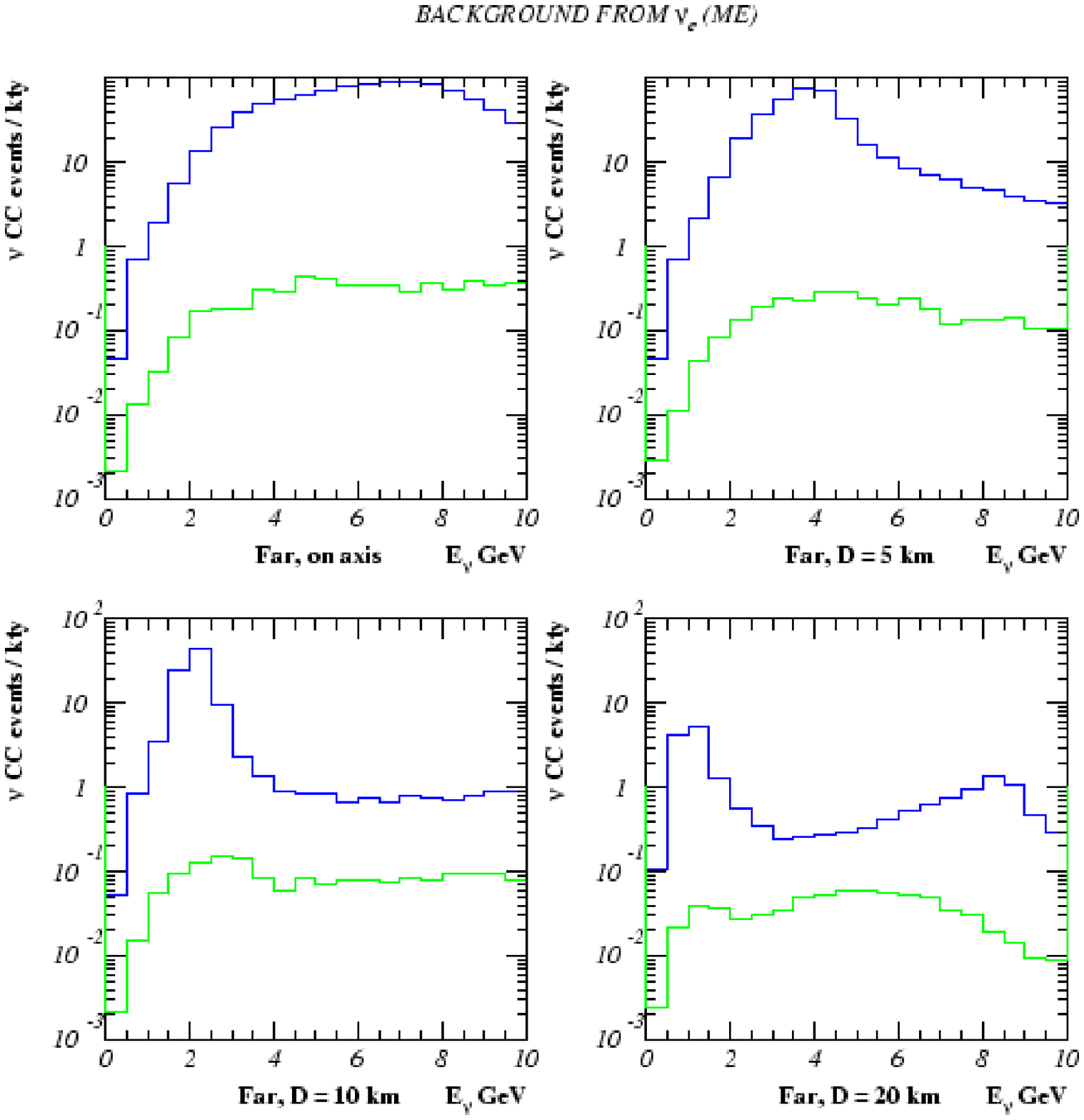}}
\caption{$\nu_{\mu}$ and $\nu_{e}$ induced CC event rates at the
on- and off-axis detectors, medium energy beam. The upper histogram represents the
$\nu_{\mu}$ rates, the lower one shows the $\nu_{e}$ rates.}
\label{nue_back_me}
\end{figure}

A relative fluxes of the $\nu_{\mu}$ and $\nu_{e}$ neutrinos at different
detector positions are show in Fig.\ref{nue_back_le} for the low energy
beam case and in Fig.\ref{nue_back_me} for the medium energy beam. 

The $\nu_{e}$ flux in the 'off-axis' detectors is of the order
of $0.5\%$ of the $\nu_{\mu}$ flux and this background is somewhat reduced 
in comparison with the 'on-axis' detector.

\section{$\nu_{e}$ Appearance Experiment}

The primary goal of the $\nu_{e}$ appearance experiments is a measurement of the 
$U_{e3}$ element of the MNS neutrino mixing matrix or the mixing angle 
$sin\theta_{13}$. The experiment is expected to detect  a small number
of $\nu_{e}$  CC interactions among the neutrino interactions registered at the far
detector. In the following we assume that the experiment is carried out at
such a combination of $L$ and $E_{\nu}$ that the oscillations generated by
$\Delta m^{2}_{23}$ dominate and that the value of the $\Delta m^{2}_{23}$
is measured precisely in the disappearance experiment.
 A number of the $\nu_{e}$  interactions is proportional to the neutrino
mixing parameters via
 
\begin{equation}
N^{appear}_{\nu_{e}} =  sin^{2}2\theta_{e\mu}\int dE \frac{dN_{\nu_{\mu}}}{dE}
sin^{2}\frac{1.27\Delta m^{2}_{23}L}{E}=sin^{2}2\theta_{e\mu}N^{disappear}_{\nu_{\mu}}
\label{mixing_angle}
\end{equation}

where

\begin{equation}
sin^{2}2\theta_{e\mu}=\frac{1}{2}sin^{2}2\theta_{13}=2\left|U_{e3}\right|^{2}
\end{equation}

 The $\nu_{e}$  CC interactions are characterized by an absence of a long, penetrating muon 
track and by a presence of an electromagnetic energy cluster due to the outgoing electron.
There are several sources of backgrounds yielding events with  very similar characteristics:
\begin{itemize}
\item high-y $\nu_{\mu}$ CC interactions with a significant fraction of the hadronic energy
produced in a form of $\pi^{0}$'s.
\item NC interactions with a significant fraction of the hadronic energy
produced in a form of $\pi^{0}$'s
\item CC interaction of $\nu_{\tau}$'s (resulting from $\nu_{\mu} \rightarrow \nu_{\tau}$ 
oscillations) with a subsequent decay $\tau \rightarrow e$. This background is particularly
troublesome if the mixing angle $sin^{2}2\theta_{e\mu}$ is much smaller than  
$sin^{2}2\theta_{\tau\mu}$.

\end{itemize}

An additional, irreducible,  contribution to the sample of potential
$\nu_{e}$  CC interactions is produced by an intrinsic $\nu_{e}$ component
of the neutrino beam.

A sensitivity of the measurement of the mixing parameters will be limited by
the statistical fluctuations of the background event sample, even if the
overall normalization of the background is known. It is, therefore, very 
important to minimize the size of the background sample, while the sample of
detected  $\nu_{e}$ CC  interactions due to the oscillations is maximized.

The optimization procedure involves several factors:
\begin{itemize}
\item a judicious choice of the beam energy and spectrum. An optimal neutrino
beam would have a large  flux concentrated around $E_{opt}=\frac{2.54L\Delta m^{2}_{23}}
{\pi}$. Such a beam would have the following effects:
   \begin{itemize}
\item  a number of $\nu_{\mu}$'s which have 'oscillated' away,
$N^{disappear}_{\nu_{\mu}}$, in Eq.\ref{mixing_angle} would be maximized
\item   background due to high-y $\nu_{\mu}$ CC  interactions would be minimized 
\item background from the NC interactions would be minimized if the neutrino spectrum
does not extend beyond the region where the oscillation probability is very large  
\end{itemize}
\item a choice of the distance of the detector from the neutrino source. To eliminate
the background due to   CC $\nu_{\tau}$ interactions it is desirable that the optimal
energy   $E_{opt}$ is below the kinematical threshold for $\tau$ production.
\item a suitable choice of the analysis cuts. Oscillations-induced $\nu_{e}$ events
will be concentrated in a narrow energy range around   $E_{opt}$. A large fraction
of NC events will have an observed energy much lower than the neutrino energy
due to the escaping neutrino, hence a requirement that the detected energy of the event 
exceeds a certain fraction of the $E_{opt}$   may significantly reduce the NC background.
The intrinsic   $\nu_{e}$ component of the beam tends to have much harder energy
spectrum (see Figs. \ref{nue_back_le} and \ref{nue_back_me}), hence a 
requirement that the observable energy does not exceed a certain value will 
significantly reduce this background.\
\end{itemize}

\begin{figure}[h]
\centerline{\includegraphics*[bb=50 180 500 640,width=12cm]{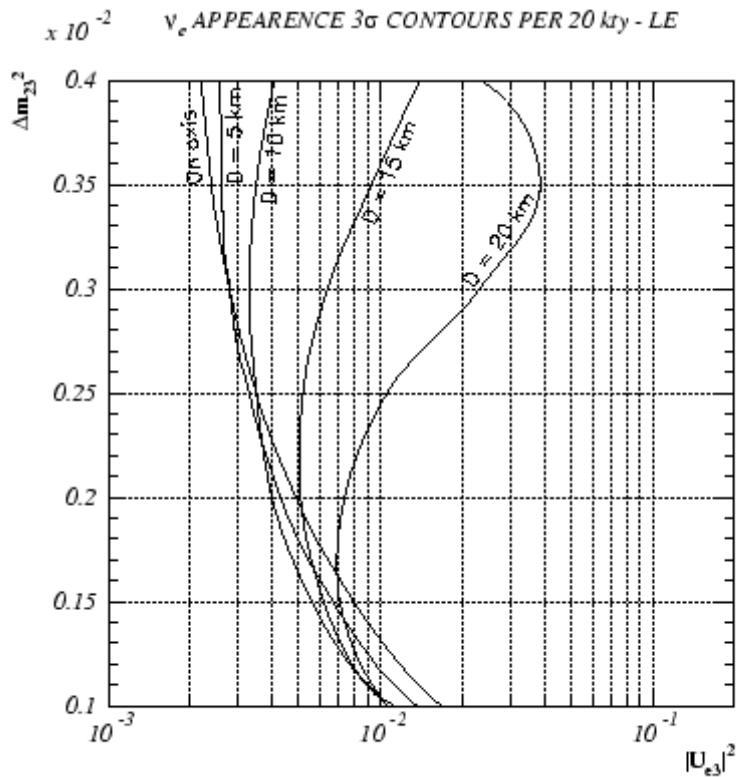}}
\vskip -0.0cm
\caption{Sensitivity of the appearance experiments to $\left|U_{e3}\right|^{2}_{min}$ 
for low energy beam exposure 20 kt*years.
}
\label{limit_le}
\end{figure}

\begin{figure}[h]
\centerline{\includegraphics*[bb=50 180 500 640,width=12cm]{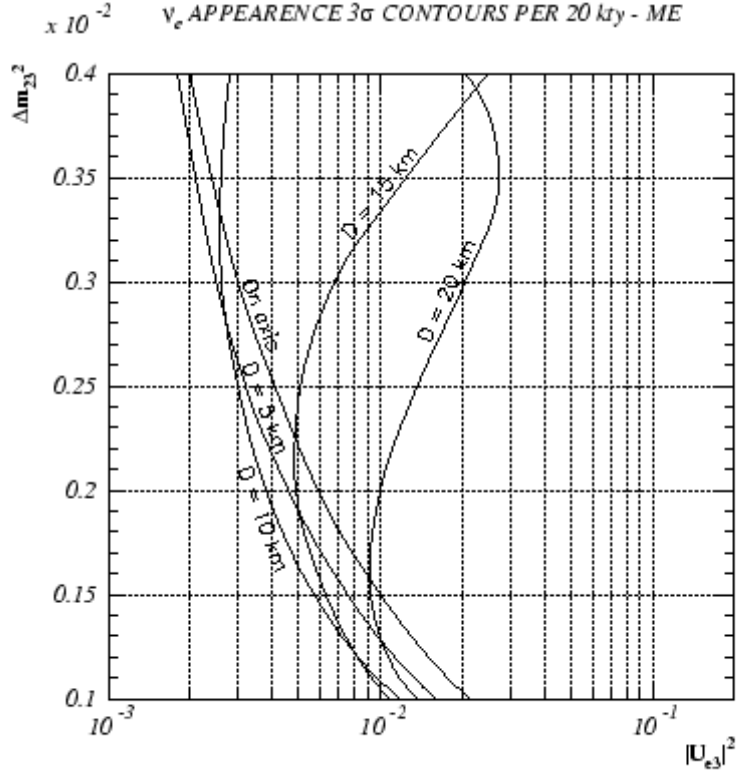}}
\caption{Sensitivity of the appearance experiments to $\left|U_{e3}\right|^{2}_{min}$ 
for medium energy beam exposure of 20 kt*years.
}
\label{limit_me}
\end{figure}

A potential of the $\nu_e$ appearance experiment with the
'off-axis' detectors is shown in Figs. \ref{limit_le} and \ref{limit_me} for the
case of low and medium energy beams respectively. Sensitivity of the experiment 
to the $\left|U_{e3}\right|^{2}$ is taken to be equal to the  3 $\sigma$ fluctuation 
of the background. It is assumed that the $\pi^{0}$ rejection power of the neutrino
detector is sufficient to reduce the NC background to the level below the intrinsic
$\nu_{e}$ component of the beam. The size of the exposure is taken to correspond
to 20 kton*years at the 100\% electron identification efficiency. 

A detector situated at the distance of 10 km from the beam axis  offers the best
opportunity for the measurement of the $\left|U_{e3}\right|^{2}$. Although a potential
reach of the detector located at the distance of 5 km is somewhat better for larger
values of $\Delta m^{2}_{32}$, the NC background would be much larger there, thus
necessitating a substantially better detector to ensure the adequate background rejection.

An example of the actual event rates for the case of medium energy beam and oscillations
corresponding to $\Delta m^{2}_{23}=0.003\:eV^{2}$ is shown in Table \ref{events}.

\begin{table}
\begin{center}
\caption{Examples of $\nu_{e}$ appearance experiments for $\Delta m^{2}_{23}=0.003\:eV^{2}$.
$E_{min}-E_{max}$ denote the optimal energy range,  $\nu_{\mu}$ no osc./osc. are the 
expected numbers of  $\nu_{\mu}$ CC interactions without and with oscillations, $\left|U_{e3}\right|^{2}_{min}$ corresponds to the value of the mixing parameters producing
an effect equal to 3 $\sigma$ fluctuation of the background.
 }
\vskip 0.5cm
\begin{tabular}
[c]{||c|c|r|r|r|r||}\hline\hline
Det. position, R & $E_{min}-E_{max}$, GeV  & $\nu_{\mu}$ no osc. & $\nu_{e}$ bckg.  & $\nu_{\mu}$ osc. & 
$\left|U_{e3}\right|^{2}_{min}$      \\\hline\hline
 0 km & 1.0 - 7.5 & 13460   & 60.0 & 9380   & 0.0030\\ \hline
 5 km & 1.0 - 4.5 &  5420   & 24.0 & 2480   & 0.0025\\ \hline
10 km & 1.5 - 2.5 &  1380   &  4.4 &   80   & 0.0024\\ \hline
15 km & 1.0 - 1.5 &   320   &  1.0 &  100   & 0.0068\\ \hline
20 km & 0.5 - 1.5 &   200   &  1.2 &  120   & 0.0206\\ \hline
\end{tabular}

\label{events}
\end{center}
\end{table}

\section{Neutrino Detectors Issues}

The NuMI neutrino beam is directed towards the MINOS detector located
in the Soudan mine. Construction of underground facilities at the off-axis
locations is not practical, therefore the design of the detectors and their
locations must be optimized to enable precise  measurements despite the
presence of cosmic rays induced backgrounds.

 
Transverse distance of the detector from the beam axis determines the average
energy of the neutrino beam. Thus the optimal detector position depends on the
value of the $\Delta m^{2}$ and on the nature of the desired measurement.

For example, if $\Delta m^{2}=3\times 10^{-3}\,eV^{2}$ then a detector 
at the transverse distance of $10\,km$ from Soudan would be located at the
maximum of the oscillation and it would be an optimal one
for the measurement of the mixing angle and for the detection of 
$\nu_{e}$ appearance. An additional detector at the distance of $20\,km$
could be used to demonstrate the oscillation pattern.

Optimization of a  possible detector location may include the variation
of the distance of the putative detectors from Fermilab. Sensitivity
of the neutrino oscillations appearance experiments does not depend on
the experiment baseline as long as the beam energy is much higher than
the one corresponding to the oscillation maximum, as $\frac{1}{L^{2}}$
reduction of the flux is compensated by the increased oscillation probability
$\left( \frac{L}{E} \right)^{2}$. 

\begin{figure}[h]
\centerline{\epsfig{file=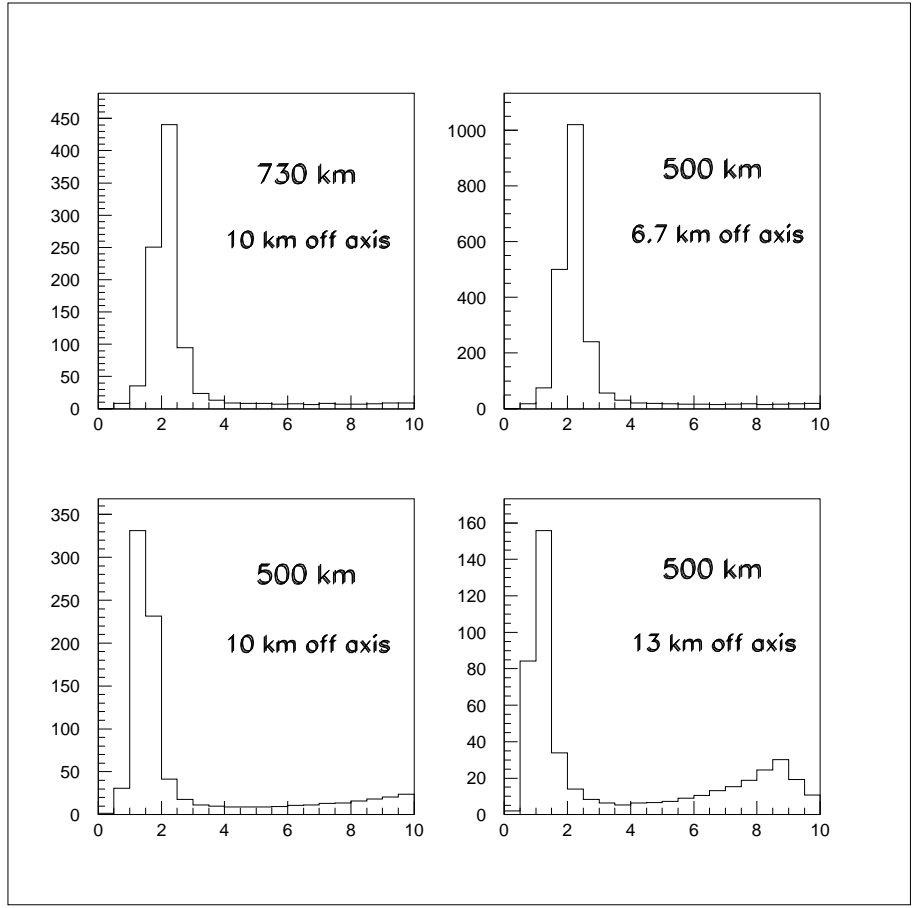,width=12cm}}
\caption{Rates of $\nu_{\mu}$  CC event rates in the detectors located
at different locations.  Rates correspond to the medium energy beam, $10\,kton\times year$ exposure}
\label{change_baseline}
\end{figure}

The situation is different when the 
combination of the experiment baseline $L$ and the beam energy $E$ corresponds
to the oscillation maximum. 
The change of the sensitivity of the appearance experiment 
with  the reduction of the baseline of the experiment is a combination of  an increase of the 
neutrino flux as $\frac{1}{L^{2}}$   and a  reduction of the oscillation 
probability, if the
detector is located at the same angle with the respect to the neutrino
axis or a change of the neutrino flux with the increased angle if the detector
is located at a larger angle to lower the neutrino energy to correspond to the oscillation maximum at the new  location. This interplay between the detector 
longitudinal and transverse positions is illustrated in Fig.\ref{change_baseline}.

The NuMI neutrino beam has an interesting property of passing, at the distance
of about $600\,km$, under the large body of water, the Lake Superior. 
It allows for an unique neutrino oscillation experiment utilizing the
same detector exposed to the narrow-band  neutrino beam of different energies.
Such an experiment mounted on a barge floating the the Lake Superior
can be  moved to an arbitrary position, thus permitting the optimization
of the neutrino beam as the understanding of the neutrino oscillation
progresses. An experiment can be also built as  submerged module(s) towed
by the barges to the requested positions. Such a scenario can significantly
 reduce the cosmic muon background. 
\section{Acknowledgments}

Numerous discussions with our collegues from NuMI/MINOS have helped us to understand
the issues presented in this paper.

 This work was supported in part by grants from the Illinois Board of
 Higher Education, the Illinois Department of Commerce and Community
 Affairs, the National Science Foundation, and the U.S. Department of
 Energy.

\end{document}